\documentclass[twocolumn,superscriptaddress,amsmath,amssymb,nofootinbib,aps,prd]{revtex4-1}
\usepackage{dcolumn}
\usepackage{bm}
\usepackage{graphicx}
\usepackage{color}

\usepackage{amsthm}

\newcommand{\be}{\begin{equation}}
\newcommand{\ee}{\end{equation}}
\newcommand{\ba}{\begin{eqnarray}}
\newcommand{\ea}{\end{eqnarray}}

\newcommand{\n}[1]{\label{#1}}
\newcommand{\eq}[1]{(\ref{#1})}
\newcommand{\hh}{\, ,\hspace{0.25cm}}

\newcommand{\ind}[1]{\mbox{\tiny #1}}

\begin{document}

\title{Testing GR with the Gravitational Wave Inspiral Signal GW170817}
\author{Andrey A. Shoom}
\email{andrey.shoom@aei.mpg.de}
\affiliation{Max Planck Institute for Gravitational Physics (Albert Einstein Institute), Leibniz Universität Hannover, Callinstr. 38, D-30167, Hannover, Germany}
\author{Pawan K. Gupta}
\email{pawan.nf@gmail.com}
\affiliation{Nikhef, Science Park, 1098 XG Amsterdam, The Netherlands}
\affiliation{Institute for Gravitational and Subatomic Physics (GRASP),\\ Department of Physics, Utrecht University, Princetonplein 1, 3584 CC Utrecht, The Netherlands}
\author{Badri Krishnan}
\email{badri.krishnan@aei.mpg.de}
\affiliation{Max Planck Institute for Gravitational Physics (Albert Einstein Institute), Leibniz Universität Hannover, Callinstr. 38, D-30167, Hannover, Germany}
\affiliation{Institute for Mathematics, Astrophysics and Particle Physics, Radboud University, Heyendaalseweg 135, 6525 AJ Nijmegen, The Netherlands}
\author{Alex B. Nielsen}
\email{alex.b.nielsen@uis.no}
\affiliation{Department of Mathematics and Physics, University of Stavanger, NO-4036 Stavanger, Norway}
\affiliation{Max Planck Institute for Gravitational Physics (Albert Einstein Institute), Leibniz Universität Hannover, Callinstr. 38, D-30167, Hannover, Germany}
\author{Collin D. Capano}
\email{collin.capano@aei.mpg.de}
\affiliation{Max Planck Institute for Gravitational Physics (Albert Einstein Institute), Leibniz Universität Hannover, Callinstr. 38, D-30167, Hannover, Germany}


\begin{abstract}

  Observations of gravitational waves from compact binary mergers have
  enabled unique tests of general relativity in the dynamical and non-linear
  regimes. One of the most important such tests are constraints on the
  post-Newtonian (PN) corrections to the phase of the
  gravitational wave signal.  The values of these PN coefficients can
  be calculated within standard general relativity, and these values
  are different in many alternate theories of gravity.
  It is clearly of great interest to constrain these deviations based
  on gravitational wave observations.  In the majority of such tests
  which have been carried out, and which yield by far the most
  stringent constraints, it is common to vary these PN coefficients
  individually.  While this might in principle be useful for detecting
  certain deviations from standard general relativity, it is a serious
  limitation.  For example, we would expect alternate theories of
  gravity to generically have additional parameters. The corrections
  to the PN coefficients would be expected to depend on these
  additional non-GR parameters whence, we expect that the various PN
  coefficients to be highly correlated.  We present an alternate
  analysis here using data from the binary neutron star coalescence
  GW170817.  Our analysis uses an appropriate linear combination of
  non-GR parameters that represent absolute deviations from the
  corresponding post-Newtonian inspiral coefficients in the TaylorF2
  approximant phase. These combinations represent uncorrelated non-GR
  parameters which correspond to principal directions of their
  covariance matrix in the parameter subspace.  Our results illustrate
  good agreement with GR. In particular, the integral non-GR phase is $\Psi_{\ind{non-GR}} = (0.447\pm253)\times10^{-1}$ and the deviation from GR percentile is $p^{\ind{Dev-GR}}_{n}=25.85\%$.

\end{abstract}

\maketitle

\section{Introduction}

The problem of binary motion in a gravitational system is one of the
oldest in astronomy. Newton's explanation of Kepler's laws was a
milestone of science. Relativistic corrections to Keplerian motion
under Newton's laws have been studied since the time of Einstein
\cite{10.2307/1968714}. These corrections can be calculated within the
post-Newtonian (PN) formalism as an expansion in the ratio $v/c$ of
the typical relative velocity $v$ between the two components of the
binary to the speed of light $c$. Low order corrections can be
observed in the motion of binary pulsar systems where typical values
of $v/c$ are $\mathcal{O}(10^{-4})$ \cite{Wex:2014nva}. The recent
observations of coalescing compact binary systems (see
e.g. \cite{Abbott:2016blz,LIGOScientific:2018mvr,TheLIGOScientific:2016pea,Nitz:2018imz,Nitz:2019hdf,Venumadhav:2019lyq,Zackay:2019tzo})
have much higher values of
$v/c \sim \mathcal{O}(10^{-1})$.  These observations therefore probe regimes where the gravitational field is strong and also
dynamical.

The relation between the binary motion and the observed gravitational
waves was first expounded by Peters and Mathews
\cite{Peters:1963ux}. This work obtained the gravitational wave phase
by calculating the Einstein quadrupole formula for underlying
Keplerian motion. Whilst this analysis gives a very approximate fit
for LIGO data \cite{Abbott:2016bqf} it has been known for a long time
that more accurate fits require a large number of relativistic
corrections \cite{Cutler:1994ys}.  Several different approaches have
been able to model the expected signal for the inspiral and
coalescence of compact objects with sufficient accuracy within
standard general relativity.  These include the Effective-one-body
models (see
e.g. \cite{Buonanno:1998gg,Damour:2016bks,Ossokine:2020kjp,Bohe:2016gbl,Taracchini:2013rva,Pan:2013rra,Pan:2011gk,Nagar:2018zoe,Damour:2014sva}),
the Phenomenological models (see
e.g. \cite{Ajith:2007kx,Khan:2019kot,Khan:2018fmp,London:2017bcn,Khan:2015jqa,Husa:2015iqa,Santamaria:2010yb})
and the surrogate models
\cite{Field:2013cfa,Blackman:2017pcm,PhysRevResearch.1.033015}.  Some
corrections have also been calculated in modified gravity theories
\cite{Ohashi:1996uz,Nielsen:2017lpd}.  In this paper we restrict our
attention to the binary neutron star merger GW170817 event
\cite{TheLIGOScientific:2017qsa}. For this event, the contribution of
the merger to the signal-to-noise-ratio is not significant, whence the
inspiral regime and standard PN methods dominate.  

It has now become routine to use the observed gravitational wave
observations to constrain relativistic corrections to binary motion
\cite{,Arun:2006yw,Arun:2006hn,Mishra:2010tp,TheLIGOScientific:2016src,TheLIGOScientific:2016pea,Abbott:2018lct,LIGOScientific:2019fpa}. The
TIGER framework \cite{Agathos:2013upa} looks for a deviation from the
calculated GR value by introducing a number of phenomenological
deviation parameters that take the value zero in GR. In this approach
the deviation parameters are assumed to be independent of the physical
parameters of the binary, such as mass and spin.

These phenomenological non-GR parameters implicitly represent a
certain class of GR alternatives.  In different GR alternatives, these
parameters may have different physical meanings. For instance, in
phenomenological modifications of GR based on the massive graviton
assumption, such that the effective Newtonian potential is of Yukawa
type with a non-standard graviton dispersion relation, the non-GR
parameter of 1PN order is proportional to the graviton's mass squared
\cite{Will:1997bb}. In other alternative theories of gravity,
radiative multipole moments differ from those of GR and their
deviation from GR values contributes to every PN order phase term
\cite{Kastha:2018bcr,Kastha:2019brk}. In general, alternatives to GR
may naturally depend on an additional set of parameters such that the
non-GR parameters are in general functionally related. This motivates
testing GR via the simultaneous variation of such non-GR parameters
introduced as additional terms in the GW phase.
    
In the tests of
\cite{TheLIGOScientific:2016pea,Abbott:2018lct,LIGOScientific:2019fpa}
two types of variations were considered: each of these parameters was
varied individually whilst keeping the others fixed, or all the
parameters were varied simultaneously. The results showed that
simultaneous variation produced wide and non-informative posteriors on
the value of each individual deviation parameter, closely consistent
with the prior \cite{TheLIGOScientific:2016src}. Whilst varying each
one individually captures a generic deviation from GR if it is strong
enough, a generic modified gravity theory is expected to differ from
GR for several of the non-GR parameters at different PN orders. This
also means that one cannot directly compare the results obtained on
each parameter individually with a theoretical calculation in an
alternate theory of gravity.  In addition, the non-GR parameters are
correlated with the PN coefficients and between themselves, which can be seen for example in Figure 7 of \cite{TheLIGOScientific:2016src}. To remedy
these limitations, the method of principal component analysis was
proposed in \cite{Ohme:2013nsa}. A similar method, known as the
singular value decomposition approach, was proposed in
\cite{Pai:2012mv}. These methods are applied to the Fisher information
matrix that estimates parameter values. However, Fisher-matrix-based
estimates are not reliable at low SNRs (see, e.g. \cite{Cutler:1994ys}
and also \cite{Vallisneri:2007ev}). This limits the measurability of
the higher-order PN terms, as well as their non-GR contributions.

In this work, in order to statistically isolate non-GR parameters we diagonalise their covariance matrix derived from the sample data arrays from the PyCBC inference package \cite{Biwer:2018osg}. This approach does not rely on the high SNR values that are required for the Fisher information matrix estimate. Using the eigenvectors of the covariance matrix we construct new uncorrelated non-GR parameters (that are linear combinations of the original ones) corresponding to the principal directions of the covariance matrix. In this work, we consider simultaneous variation of non-GR parameters introduced additively to the PN phase expansion parameters from 0PN to 2PN order, although it could be extended to other orders. We demonstrate our method on the signal GW170817 that was identified by the LIGO-Virgo detector network as the first gravitational wave observation of a binary neutron star inspiral \cite{TheLIGOScientific:2017qsa}.

Although, the combined signal-to-noise ratio of 32.4 for GW170817 was the largest of any gravitational wave event observed to date, our primary interest in this system is because of its low mass and long inspiral phase in the detector band. The chirp mass was found to be ${\cal M}=1.188^{+0.004}_{-0.002}\,M_{\odot}$ and with the restriction of the component spins to the range inferred in binary neutron stars, the component masses were found to be in the range $1.17-1.60\,M_{\odot}$, with the total system mass of $2.74^{+0.04}_{-0.01}\,M_{\odot}$. The signal source sky position was localised to an area of 28 deg$^2$ with 90$\%$ probability and had a luminosity distance of $40^{+8}_{-14}$ Mpc.

Our paper is organised as follows. In Sec.~II we introduce the waveform model and its TaylorF2 approximation up to the 4PN order. We also introduce a set of non-GR additive parameters to the corresponding PN coefficients. All these parameters are given uniform prior distributions and they are varied simultaneously. We derive the corresponding posteriors of all the waveform parameters. Next, diagonalizing of the non-GR parameters covariance matrix, we construct new uncorrelated non-GR parameters and repeat the procedure. In Sec.~III we analyse posteriors of the non-GR parameters and define two different methods to measure deviations from GR. We summarise our results in the Conclusion.       

\section{Posterior distributions of the non-GR parameters}

When the rate of change of frequency is small relative to the squared frequency, the expansion of a GW signal $h(t)$ in the Fourier domain can be obtained from an expansion in the time domain using the Stationary Phase Approximation \cite{Cutler:1994ys}. For the dominant frequency, the Fourier domain waveform can be written as
\be\n{3.1}
\tilde{h}(f) = A(f)e^{i\Psi(f)} 
\ee
where $\tilde{h}(f)$ is the Fourier transform of $h(t)$ and 
\ba\n{3.2} 
\Psi(f) & = & \sum^{8}_{n=0}\left(\frac{f}{f_{0}}\right)^{(n-5)/3}\left[\Psi_{n}+\Psi^{\ln}_{n}\ln\left(\frac{f}{f_{0}}\right)\right]\,.
\ea
Here $\Psi_{1}\equiv0$ and the only non-zero logarithmic terms are $\Psi^{\ln}_{5}$ and $\Psi^{\ln}_{6}$. The expansion order $n$ corresponds to $(n/2)$PN term. In this expansion the 2.5PN term becomes indistinguishable from the binary coalescence phase $(-\phi_{c}-\pi/4)$ and the 4PN term is indistinguishable from the binary coalescence time $2\pi ft_{c}$. The gravitational wave amplitude $A$ can also be expanded as a post-Newtonian series, but here we will keep only the terms at Newtonian order and hence $A(f) \sim f^{-7/6}$. Factors of a reference frequency $f_{0}$ are included as in \cite{Ohme:2013nsa} to render the expansion coefficients explicitly dimensionless. In what follows, we shall take the ``natural" reference frequency $f_{\ind{ref}}=1/(\pi M)$. This choice matches the expansion \eq{3.2} of the TaylorF2 approximant form (see, e.g., \cite{Buonanno:2009zt}), which is also implemented in the LALSuite software package \cite{lalsuite}. 

\begin{figure*}[p]
\begin{center}
\includegraphics[width=18cm]{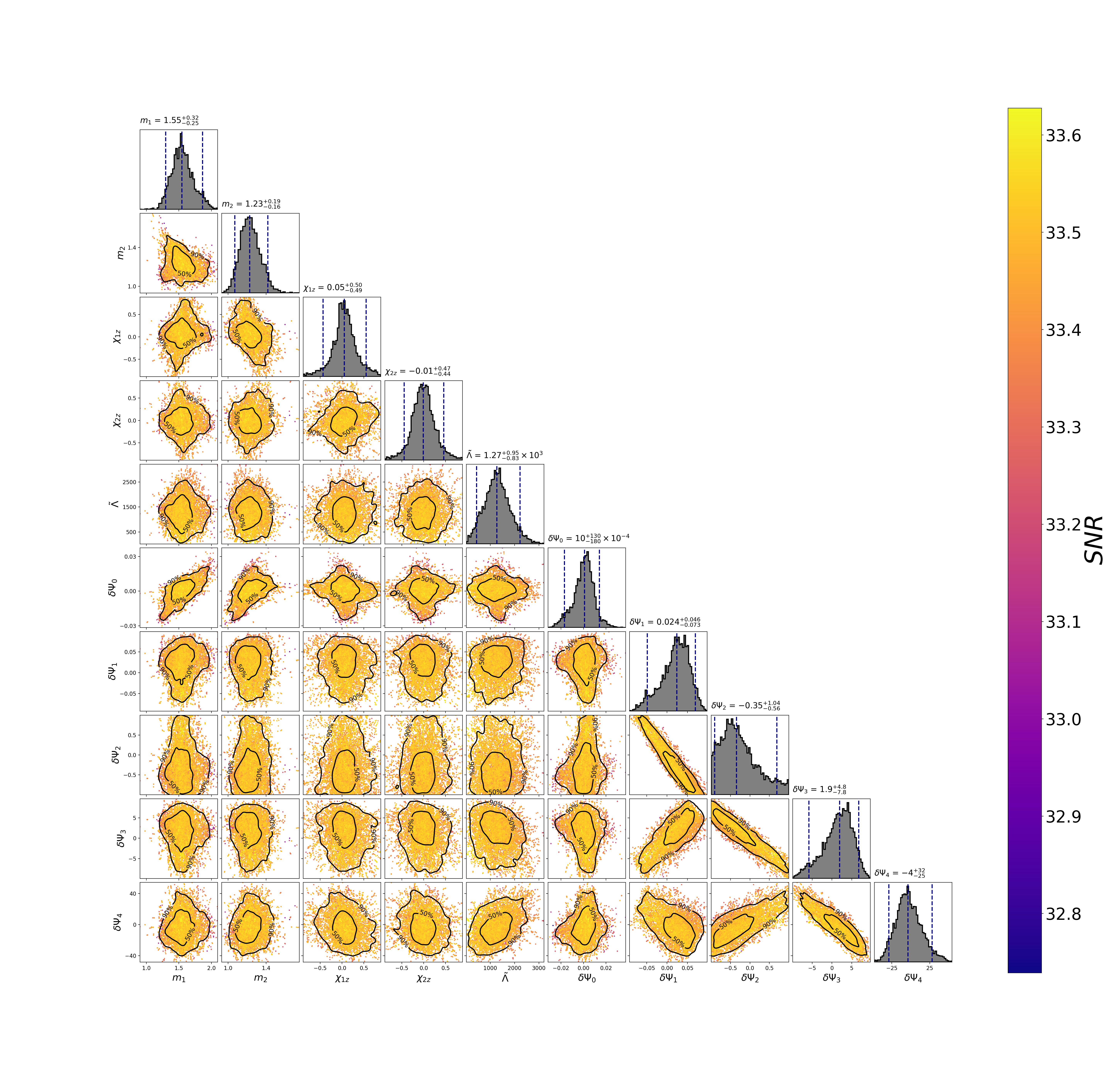}
\caption{Posterior density distributions of the binary parameters and the non-GR parameters. The central plots are 2D marginal posteriors, where the black contours show the 50\% and 90\% credible regions. The upper and the right plots are the 1D marginal posteriors, where the median and 90\% credible intervals are indicated by the dashed lines.}\label{fig1}
\end{center}
\end{figure*}

\begin{figure*}[p]
\begin{center}
\includegraphics[width=18cm]{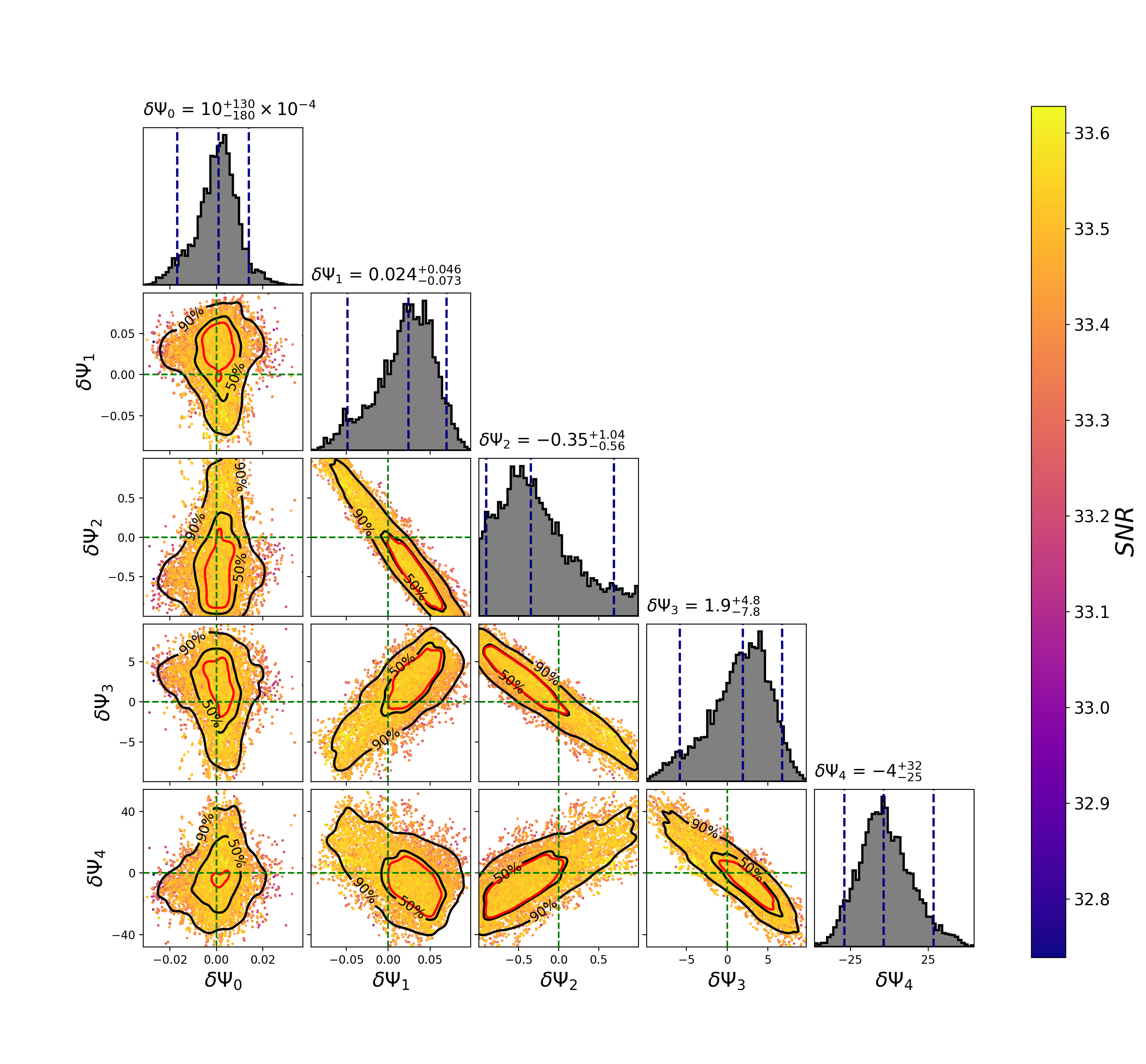}
\caption{Posterior density distributions of the non-GR parameters from Fig. 1. The central plots are 2D marginal posteriors, where the black contours show the 50\% and 90\% credible regions. The upper and the right plots are the 1D marginal posteriors, where the median and 90\% credible intervals are indicated by the dashed lines. Red contours are defined by vanishing non-GR parameters indicated by intersection of the green dashed lines. They define confidence regions of the 2D joint distributions. }\label{fig2}
\end{center}
\end{figure*}

The physical parameters that enter our waveform are the binary masses $m_1$ and $m_2$, the orbit-aligned dimensionless spin components of the stars $\chi_{z1}$ and $\chi_{z2}$ (we consider high-spin priors), the dimensionless deformability parameters $\Lambda_1$ and  $\Lambda_2$, which already appear in the 5PN term (here we present the posterior of the dimensionless combined tidal deformability $\tilde{\Lambda}$), and the following set of non-GR parameters\footnote{Here we restricted the non-GR parameters up to the 2PN order. Measurements of the higher orders are less accurate and their posteriors do not converge well. We include point-particle PN terms from GR up to 4PN and tidal deformation terms up to 6PN.}:
\be\n{3.3}
\delta\Psi_{0}\hh \delta\Psi_{1}\hh \delta\Psi_{2}\hh \delta\Psi_{3}\hh \delta\Psi_{4}\,,
\ee
that represent the {\em absolute} deviations in their corresponding PN coefficients, 
\be\n{3.4}
\Psi_{n}\longrightarrow\Psi_{n}+\delta\Psi_{n}\,.
\ee
To obtain the posterior density distributions of these parameters we exploit the PyCBC inference package \cite{Biwer:2018osg}. In particular, we use the marginalized phase model and the emcee\_pt sampler with 2000 walkers, 12000 effective samples, with a maximum of 1000 samples per chain, and four temperatures. We used (uniform) prior distributions and varied all the parameters {\em simultaneously}. One- and two-dimensional posteriors of all the parameters are shown in Fig.~\ref{fig1} and posteriors of the non-GR parameters are shown separately in Fig.~\ref{fig2}. One can notice that the dimensionless combined tidal deformability $\tilde{\Lambda}$ is correlated with other parameters, and in particular with $\delta\Psi_{4}$ non-GR parameter.

One can infer from Fig.~\ref{fig2} that the non-GR parameters are
strongly correlated, in particular $\delta\Psi_{n}$ and
$\delta\Psi_{n+1}$, for $n=1,2,3$. The off-diagonal (correlation)
terms corresponding to these parameters in the $(5\times5)$ covariance
matrix $\Sigma_{nm}$ make these parameters less accurately
measured. Moreover, because of the correlation, we cannot infer
directly the accuracy of these parameters from their 1D
posteriors. Thus, we are to define a related set of uncorrelated
non-GR parameters.\footnote{Ideally, we could construct a complete set
  of all 10 uncorrelated parameters. However, such parameters would
  represent a mixture of the physical binary parameters and the non-GR
  parameters and this would alter their original physical
  meaning. Thus, to avoid this situation, we keep the non-GR sector in
  the parameter space analytically (but not statistically) isolated.}
To construct such a set we need to diagonalize the covariance matrix
$\Sigma_{nm}$.  We solve the eigenvalue-eigenvectors problem and
derive the transformation matrix $T_{nm}$, which diagonalises the
covariance matrix. Its transposed form $T_{mn}$ relates the correlated
and new uncorrelated non-GR parameters $\delta\hat{\Psi}_{n}$:
\be\n{3.5} 
\delta\hat{\Psi}_{m}=T_{mn}\delta{\Psi}_{n}\,.
\ee
These matrix elements $T_{mn}$ will obviously depend on the detector
noise spectrum and on the source parameters as well.  Explicitly, the
matrix elements which define the transformation are:
\begin{widetext}
\be\n{3.6}
\left[\begin{array}{c}
\delta\hat{\Psi}_{0} \\
\delta\hat{\Psi}_{1} \\
\delta\hat{\Psi}_{2} \\
\delta\hat{\Psi}_{3} \\
\delta\hat{\Psi}_{4} \\
\end{array}\right] = 
\left[\begin{array}{ccccc}
 \,\,\,\,\,3.03\times10^{-1} & -9.47\times10^{-1} & -1.08\times10^{-1} & -4.00\times10^{-3} &\,\,\,\,\,4.69\times10^{-4} \\
 \,\,\,\,\,9.53\times10^{-1} & \,\,\,\,\,3.02\times10^{-1} &\,\,\,\,\,2.98\times10^{-2} &\,\,\,\,\,5.57\times10^{-4} & -3.04\times10^{-4} \\
 \,\,\,\,\,4.24\times10^{-3} & -1.11\times10^{-1} &\,\,\,\,\,9.80\times10^{-1} &\,\,\,\,\,1.63\times10^{-1} &\,\,\,\,\,1.17\times10^{-2} \\
-1.21\times10^{-5} & -1.43\times10^{-2} &\,\,\,\,\,1.61\times10^{-1} & -9.68\times10^{-1} & -1.90\times10^{-1} \\
 \,\,\,\,\,9.73\times10^{-5} & -9.11\times10^{-4} &\,\,\,\,\,1.97\times10^{-2} & -1.90\times10^{-1} &\,\,\,\,\,9.82\times10^{-1} \\
\end{array}\right]\cdot
\left[\begin{array}{c}
\delta{\Psi}_{0} \\
\delta{\Psi}_{1} \\
\delta{\Psi}_{2} \\
\delta{\Psi}_{3} \\
\delta{\Psi}_{4} \\
\end{array}\right]
\ee
\end{widetext}
The eigenvalue-eigenvectors problem computations were performed using
the Householder reduction followed by the QL algorithm. Details of these methods can be found
in the book \cite{Fortran}.

\begin{figure*}[p]
\begin{center}
\includegraphics[width=18cm]{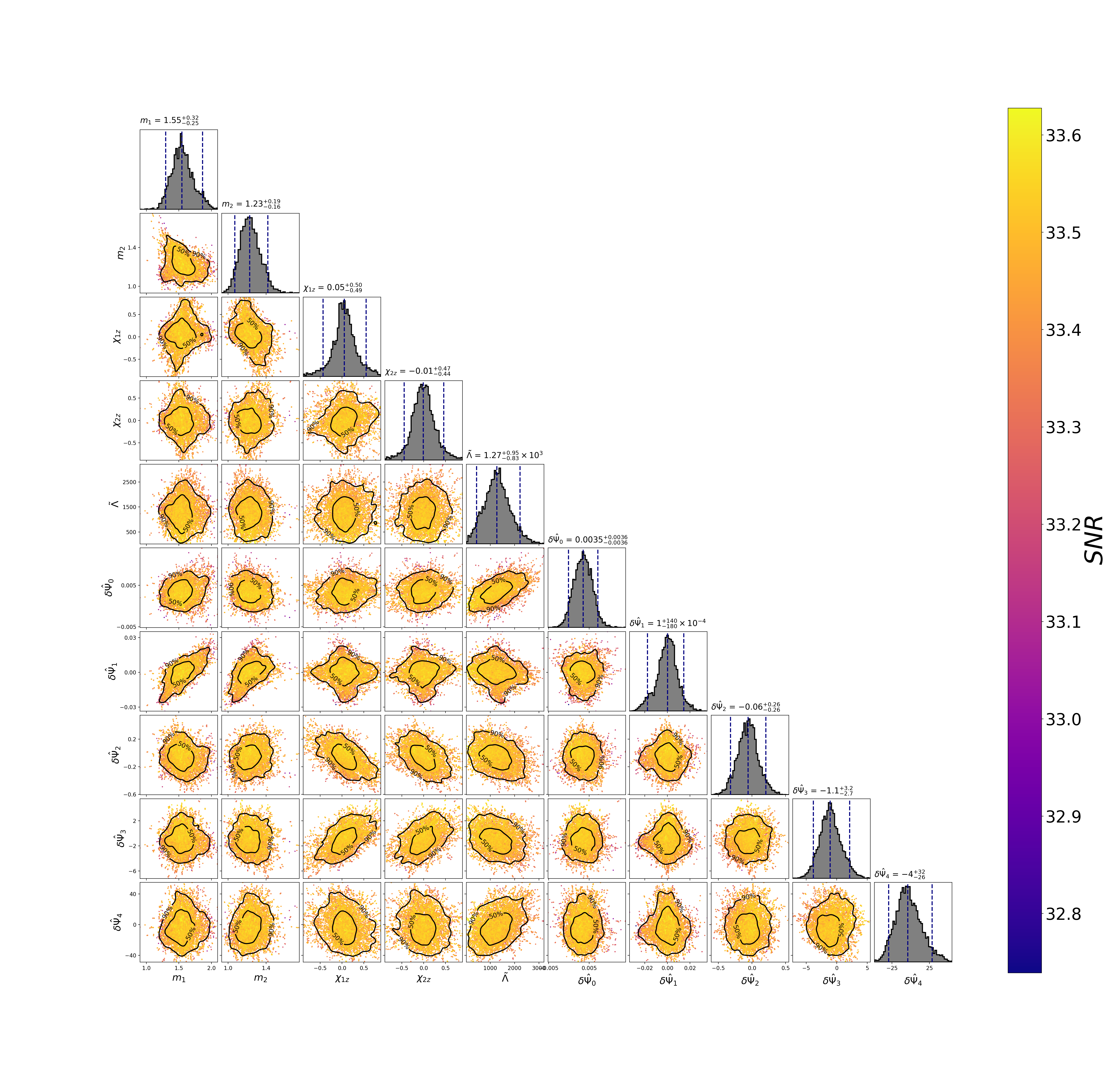}
\caption{Posterior density distributions of the binary parameters and the uncorrelated non-GR parameters. The central plots are 2D marginal posteriors, where the black contours show the 50\% and 90\% credible regions. The upper and the right plots are the 1D marginal posteriors, where the median and 90\% credible intervals are indicated by the dashed lines.}\label{fig3}
\end{center}
\end{figure*}

\begin{figure*}[p]
\begin{center}
\includegraphics[width=18cm]{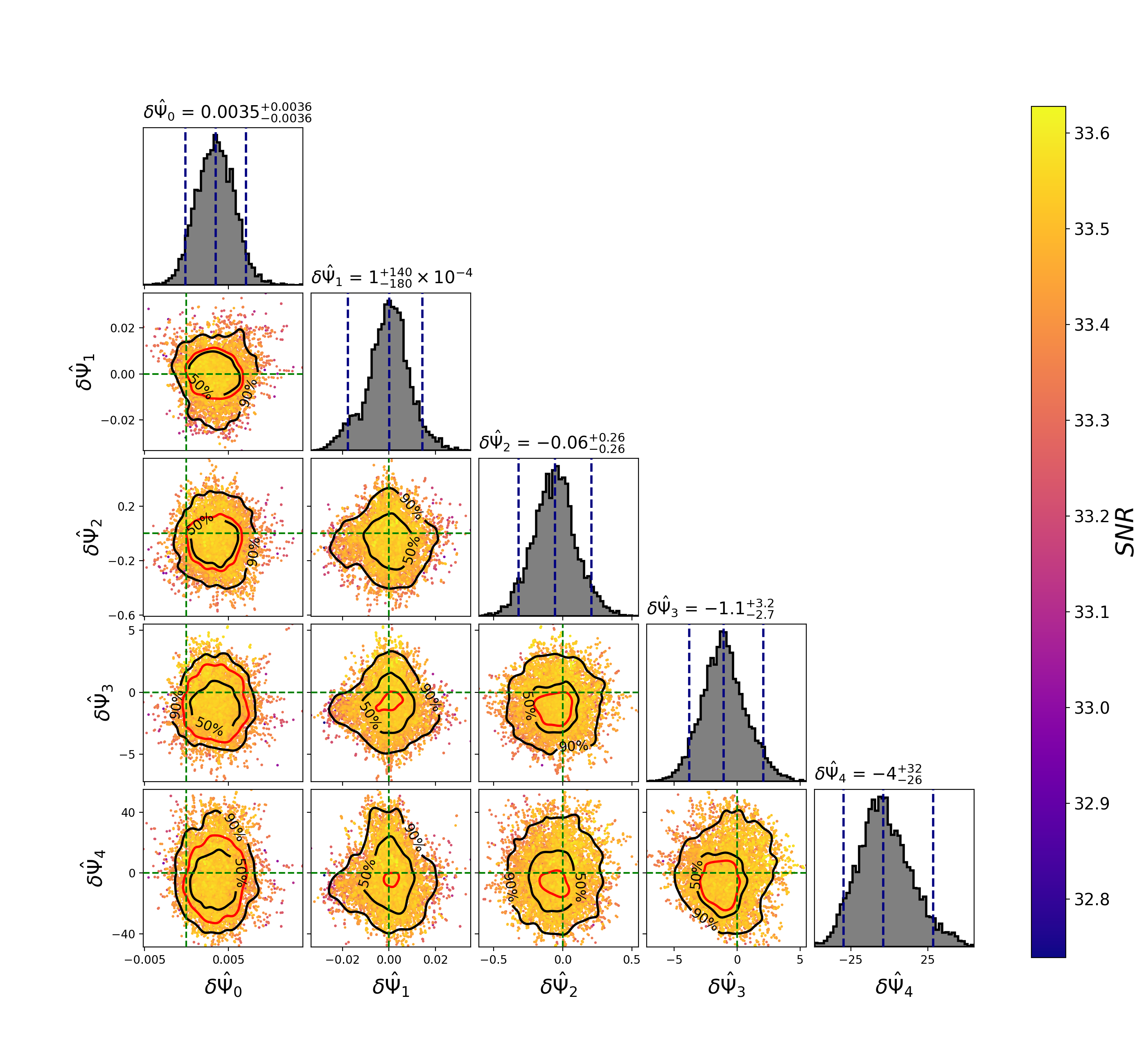}
\caption{Posterior density distributions of the uncorrelated non-GR parameters. The central plots are 2D marginal posteriors, where the black contours show the 50\% and 90\% credible regions. The upper and the right plots are the 1D marginal posteriors, where the median and 90\% credible intervals are indicated by the dashed lines. Red contours are defined by vanishing non-GR parameters indicated by intersection of the green dashed lines.. They define confidence regions of the 2D joint distributions. For the $\delta\hat{\Psi}_{1}$ and $\delta\hat{\Psi}_{2}$ joint distribution, the GR confidence region is degenerated to the single point $(0,0)$.}\label{fig4}
\end{center}
\end{figure*}

To illustrate the result of this transformation we present one- and two-dimensional posteriors of the binary parameters together with the uncorrelated non-GR parameters in Fig.~\ref{fig3} and separately the posteriors of the non-GR parameters in Fig.~\ref{fig4}. 

One can see that the new non-GR parameters are indeed almost
uncorrelated. Moreover, the measured values of the new non-GR
parameters are overall more constrained, i.e., more accurate. We note
that the presence of the non-GR parameters in the waveform model does
affect the values of the physical parameters corresponding to GR
waveform model. However, the 90\% credible intervals of the ``new"
values of physical parameters and the ``old" ones largely overlap.

\section{Analysis of the results: Measures of deviations from GR}

From the posteriors of the non-GR parameters one can immediately infer that the GR prediction, i.e. $\delta\Psi_{n}=0$, $n=0,...,4$ falls well within their 90\% credible intervals. However, one may prefer to have additional quantitative estimates of GR validity. In this section, we measure deviations from GR based on our derived results. 

\subsection{Integral non-GR phase}

According to our approach \eq{3.2}, \eq{3.4}, the waveform phase is an additive combination of the GR and non-GR components,
\be\n{4.1}
\Psi(f)=\Psi_{\ind{GR}}(f)+\Psi_{\ind{non-GR}}(f)\,,
\ee
where
\be\n{4.2} 
\Psi_{\ind{non-GR}}(f) = \sum^{4}_{n=0}\left(\frac{f}{f_{\ind{ref}}}\right)^{(n-5)/3}\delta\Psi_{n}\,
\ee
 is the non-GR part of the waveform phase. In order to measure the non-GR component value we introduce the {\em integral non-GR phase}
\be\n{4.3}
\Psi_{\ind{non-GR}}=\int_{f_{\ind{min}}/f_{\ind{ref}}}^{f_{\ind{max}}/f_{\ind{ref}}}\Psi_{\ind{non-GR}}(f)d\left(f/f_{\ind{ref}}\right)\,.
\ee
The integral non-GR phase can be computed by using the estimated values of the non-GR parameters [see Fig.~\ref{fig2}] and the expression \eq{4.2} above. 

We may also estimate uncertainty $\sigma$ of the non-GR phase by using the propagation of errors approach. The variance of the function \eq{4.2} is 
\be\n{4.4} 
\sigma^2(f) = \sum^{4}_{n,m=0}\left(\frac{f}{f_{\ind{ref}}}\right)^{(m-5)/3}\Sigma_{nm}\left(\frac{f}{f_{\ind{ref}}}\right)^{(n-5)/3}\,.
\ee
Integrating this expression in the frequency range, we derive the variance of the non-GR phase
\be\n{4.5}
\sigma^2=\int_{f_{\ind{min}}/f_{\ind{ref}}}^{f_{\ind{max}}/f_{\ind{ref}}}\sigma^2(f)d\left(f/f_{\ind{ref}}\right)\,.
\ee
For our data we find
\be\n{4.6}
\Psi_{\ind{non-GR}} = (0.447\pm253)\times10^{-1}\,.
\ee
Thus, the GR prediction $\Psi_{\ind{non-GR}}=0$ falls well within \eq{4.6}.

\subsection{Confidence regions and deviation from GR percentile}

Here we present another ways to measure GR validity. First, we construct confidence regions whose boundary corresponds to the predicted by GR zero values of the non-GR parameters. This approach is based on the kernel density estimation (kde). For a 2D kde of each pair of the non-GR parameters we used the Python function scipy.stats.gaussian\_kde from the SciPy library, that is implemented by PyCBC by the default. The confidence regions are shown in Fig.~\ref{fig2} and Fig.~\ref{fig4}.

Second, we calculate the deviation from GR percentile $p^{\ind{Dev-GR}}_{n}$ in the following way. The PyCBC inference hdf file contains the posterior for original non-GR PN parameters. We extract the effective samples from the inference hdf file by using pycbc\_inference\_extract\_samples command to get a new hdf file. From this new hdf file, we read the posterior for each original non-GR PN parameters $\delta\Psi_n$'s as an array using the standard reading hdf file command. In order to get array for each new non-GR parameter $\delta\hat{\Psi}_{n}$, we use the expression \eq{3.6}. This gives us array for each new non-GR parameter $\delta\hat{\Psi}_{n}$. To get the probability density $p(\delta\hat{\Psi}_{n})$ we use the Python function scipy.stats.gaussian\_kde.

The original non-GR PN parameters are correlated, but the new non-GR parameters are almost not correlated. Exploiting this fact, we construct the probability density $P$ in 5 dimensions for the new non-GR parameters by multiplying the individual probability densities $p(\delta\hat{\Psi}_{n})$'s of the new non-GR parameters,
\be\n{4.7}
P(\delta\hat{\Psi}_{n};n=0,...,4) = \prod_{n=0}^{4}  p(\delta\hat{\Psi}_{n})\,.
\ee
Next, we define the level hypersurface $\Sigma:\,P=\text{const}$ corresponding to GR prediction, i.e. to zero values of the new non-GR parameters,
\be\n{4.8}
\Sigma:\,P(\delta\hat{\Psi}_{n};n=0,...,4) = P(\delta\hat{\Psi}_n=0;n=0,...,4)\,.
\ee
Finally, to get the deviation from GR percentile, we integrate $P$ over the domain $D(\Sigma)$ enclosed by this hypersurface,
\be\n{4.9}
p^{\text{Dev-GR}}_{n} = \int_{D(\Sigma)} P \,d^5(\delta\hat{\Psi}_{n})\,.
\ee
To get the value 
\be\n{4.10}
p^{\ind{Dev-GR}}_{n}=25.85\%
\ee
of the deviation from GR percentile, we performed the integration numerically by using the vegas package, which exploits the adaptive multidimensional Monte Carlo integration.

\section{Conclusion} 

We have analysed deviations from GR with the example of the binary neutron star signal GW170817. A set of 5 non-GR parameters up to and including the 2PN order was introduced into the GW waveform in the TaylorF2 phase approximant. These parameters represent {\em absolute} deviations of the post-Newtonian coefficients from their GR values. By using the PyCBC package we computed their posterior density distributions for the simultaneous variation of all these non-GR parameters using uniform prior distributions. Next, diagonalizing their $(5\times5)$ covariance matrix, we derived linear combinations of these parameters that represent a set of uncorrelated non-GR parameters. This approach allows the first study of each of the new non-GR parameters separately. Each of these uncorrelated parameters corresponds to a covariance matrix principal direction in the 5D subspace of the parameter space.

Our results provide more stringent constraints on GR than those presented in \cite{TheLIGOScientific:2016pea,Abbott:2018lct,LIGOScientific:2019fpa}. In comparison to the principal component analysis \cite{Ohme:2013nsa} or the singular value decomposition approach \cite{Pai:2012mv} applied to the Fisher matrix, our computations provide a more efficient technique that achieves more stringent constraints on parameters estimation. Although the approach presented here, based on orthogonalisation of the covariance matrix, looks quite promising, one can also analyse a ``hybrid method", that combines both the principal component analysis of the Fisher matrix and the orthogonalisation of the covariance matrix constructed from the computed data. A more detailed study of this possibility is left for future work.

\begin{acknowledgments}

  The computation of the work was run on the ATLAS computing cluster
  at AEI Hannover \cite{Atlas} funded by the Max Planck Society and
  the State of Niedersachsen, Germany.
  
  This research has made use of data, software and/or web tools
  obtained from the LIGO Open Science Center
  (\url{https://losc.ligo.org}), a service of LIGO Laboratory, the
  LIGO Scientific Collaboration and the Virgo Collaboration.  LIGO is
  funded by the U.S. National Science Foundation. Virgo is funded by
  the French Centre National de Recherche Scientifique (CNRS), the
  Italian Istituto Nazionale della Fisica Nucleare (INFN) and the
  Dutch Nikhef, with contributions by Polish and Hungarian institutes.

\end{acknowledgments}

\bibliography{biblio}{}

\begin{thebibliography}{52}%
\makeatletter
\providecommand \@ifxundefined [1]{%
 \@ifx{#1\undefined}
}%
\providecommand \@ifnum [1]{%
 \ifnum #1\expandafter \@firstoftwo
 \else \expandafter \@secondoftwo
 \fi
}%
\providecommand \@ifx [1]{%
 \ifx #1\expandafter \@firstoftwo
 \else \expandafter \@secondoftwo
 \fi
}%
\providecommand \natexlab [1]{#1}%
\providecommand \enquote  [1]{``#1''}%
\providecommand \bibnamefont  [1]{#1}%
\providecommand \bibfnamefont [1]{#1}%
\providecommand \citenamefont [1]{#1}%
\providecommand \href@noop [0]{\@secondoftwo}%
\providecommand \href [0]{\begingroup \@sanitize@url \@href}%
\providecommand \@href[1]{\@@startlink{#1}\@@href}%
\providecommand \@@href[1]{\endgroup#1\@@endlink}%
\providecommand \@sanitize@url [0]{\catcode `\\12\catcode `\$12\catcode
  `\&12\catcode `\#12\catcode `\^12\catcode `\_12\catcode `\%12\relax}%
\providecommand \@@startlink[1]{}%
\providecommand \@@endlink[0]{}%
\providecommand \url  [0]{\begingroup\@sanitize@url \@url }%
\providecommand \@url [1]{\endgroup\@href {#1}{\urlprefix }}%
\providecommand \urlprefix  [0]{URL }%
\providecommand \Eprint [0]{\href }%
\providecommand \doibase [0]{http://dx.doi.org/}%
\providecommand \selectlanguage [0]{\@gobble}%
\providecommand \bibinfo  [0]{\@secondoftwo}%
\providecommand \bibfield  [0]{\@secondoftwo}%
\providecommand \translation [1]{[#1]}%
\providecommand \BibitemOpen [0]{}%
\providecommand \bibitemStop [0]{}%
\providecommand \bibitemNoStop [0]{.\EOS\space}%
\providecommand \EOS [0]{\spacefactor3000\relax}%
\providecommand \BibitemShut  [1]{\csname bibitem#1\endcsname}%
\let\auto@bib@innerbib\@empty
\bibitem [{\citenamefont {Einstein}\ \emph {et~al.}(1938)\citenamefont
  {Einstein}, \citenamefont {Infeld},\ and\ \citenamefont
  {Hoffmann}}]{10.2307/1968714}%
  \BibitemOpen
  \bibfield  {author} {\bibinfo {author} {\bibfnamefont {A.}~\bibnamefont
  {Einstein}}, \bibinfo {author} {\bibfnamefont {L.}~\bibnamefont {Infeld}}, \
  and\ \bibinfo {author} {\bibfnamefont {B.}~\bibnamefont {Hoffmann}},\ }\href
  {http://www.jstor.org/stable/1968714} {\bibfield  {journal} {\bibinfo
  {journal} {Annals of Mathematics}\ }\textbf {\bibinfo {volume} {39}},\
  \bibinfo {pages} {65} (\bibinfo {year} {1938})}\BibitemShut {NoStop}%
\bibitem [{\citenamefont {Wex}(2014)}]{Wex:2014nva}%
  \BibitemOpen
  \bibfield  {author} {\bibinfo {author} {\bibfnamefont {N.}~\bibnamefont
  {Wex}},\ }\href@noop {} {\  (\bibinfo {year} {2014})},\ \Eprint
  {http://arxiv.org/abs/1402.5594} {arXiv:1402.5594 [gr-qc]} \BibitemShut
  {NoStop}%
\bibitem [{\citenamefont {Abbott}\ \emph
  {et~al.}(2016{\natexlab{a}})\citenamefont {Abbott} \emph
  {et~al.}}]{Abbott:2016blz}%
  \BibitemOpen
  \bibfield  {author} {\bibinfo {author} {\bibfnamefont {B.~P.}\ \bibnamefont
  {Abbott}} \emph {et~al.} (\bibinfo {collaboration} {LIGO Scientific,
  Virgo}),\ }\href {\doibase 10.1103/PhysRevLett.116.061102} {\bibfield
  {journal} {\bibinfo  {journal} {Phys. Rev. Lett.}\ }\textbf {\bibinfo
  {volume} {116}},\ \bibinfo {pages} {061102} (\bibinfo {year}
  {2016}{\natexlab{a}})},\ \Eprint {http://arxiv.org/abs/1602.03837}
  {arXiv:1602.03837 [gr-qc]} \BibitemShut {NoStop}%
\bibitem [{\citenamefont {Abbott}\ \emph
  {et~al.}(2019{\natexlab{a}})\citenamefont {Abbott} \emph
  {et~al.}}]{LIGOScientific:2018mvr}%
  \BibitemOpen
  \bibfield  {author} {\bibinfo {author} {\bibfnamefont {B.~P.}\ \bibnamefont
  {Abbott}} \emph {et~al.} (\bibinfo {collaboration} {LIGO Scientific,
  Virgo}),\ }\href {\doibase 10.1103/PhysRevX.9.031040} {\bibfield  {journal}
  {\bibinfo  {journal} {Phys. Rev. X}\ }\textbf {\bibinfo {volume} {9}},\
  \bibinfo {pages} {031040} (\bibinfo {year} {2019}{\natexlab{a}})},\ \Eprint
  {http://arxiv.org/abs/1811.12907} {arXiv:1811.12907 [astro-ph.HE]}
  \BibitemShut {NoStop}%
\bibitem [{\citenamefont {Abbott}\ \emph
  {et~al.}(2016{\natexlab{b}})\citenamefont {Abbott} \emph
  {et~al.}}]{TheLIGOScientific:2016pea}%
  \BibitemOpen
  \bibfield  {author} {\bibinfo {author} {\bibfnamefont {B.~P.}\ \bibnamefont
  {Abbott}} \emph {et~al.} (\bibinfo {collaboration} {LIGO Scientific,
  Virgo}),\ }\href {\doibase 10.1103/PhysRevX.6.041015} {\bibfield  {journal}
  {\bibinfo  {journal} {Phys. Rev. X}\ }\textbf {\bibinfo {volume} {6}},\
  \bibinfo {pages} {041015} (\bibinfo {year} {2016}{\natexlab{b}})},\ \bibinfo
  {note} {[Erratum: Phys.Rev.X 8, 039903 (2018)]},\ \Eprint
  {http://arxiv.org/abs/1606.04856} {arXiv:1606.04856 [gr-qc]} \BibitemShut
  {NoStop}%
\bibitem [{\citenamefont {Nitz}\ \emph {et~al.}(2019)\citenamefont {Nitz},
  \citenamefont {Capano}, \citenamefont {Nielsen}, \citenamefont {Reyes},
  \citenamefont {White}, \citenamefont {Brown},\ and\ \citenamefont
  {Krishnan}}]{Nitz:2018imz}%
  \BibitemOpen
  \bibfield  {author} {\bibinfo {author} {\bibfnamefont {A.~H.}\ \bibnamefont
  {Nitz}}, \bibinfo {author} {\bibfnamefont {C.}~\bibnamefont {Capano}},
  \bibinfo {author} {\bibfnamefont {A.~B.}\ \bibnamefont {Nielsen}}, \bibinfo
  {author} {\bibfnamefont {S.}~\bibnamefont {Reyes}}, \bibinfo {author}
  {\bibfnamefont {R.}~\bibnamefont {White}}, \bibinfo {author} {\bibfnamefont
  {D.~A.}\ \bibnamefont {Brown}}, \ and\ \bibinfo {author} {\bibfnamefont
  {B.}~\bibnamefont {Krishnan}},\ }\href {\doibase 10.3847/1538-4357/ab0108}
  {\bibfield  {journal} {\bibinfo  {journal} {Astrophys. J.}\ }\textbf
  {\bibinfo {volume} {872}},\ \bibinfo {pages} {195} (\bibinfo {year}
  {2019})},\ \Eprint {http://arxiv.org/abs/1811.01921} {arXiv:1811.01921
  [gr-qc]} \BibitemShut {NoStop}%
\bibitem [{\citenamefont {Nitz}\ \emph {et~al.}(2020)\citenamefont {Nitz},
  \citenamefont {Dent}, \citenamefont {Davies}, \citenamefont {Kumar},
  \citenamefont {Capano}, \citenamefont {Harry}, \citenamefont {Mozzon},
  \citenamefont {Nuttall}, \citenamefont {Lundgren},\ and\ \citenamefont
  {T\'apai}}]{Nitz:2019hdf}%
  \BibitemOpen
  \bibfield  {author} {\bibinfo {author} {\bibfnamefont {A.~H.}\ \bibnamefont
  {Nitz}}, \bibinfo {author} {\bibfnamefont {T.}~\bibnamefont {Dent}}, \bibinfo
  {author} {\bibfnamefont {G.~S.}\ \bibnamefont {Davies}}, \bibinfo {author}
  {\bibfnamefont {S.}~\bibnamefont {Kumar}}, \bibinfo {author} {\bibfnamefont
  {C.~D.}\ \bibnamefont {Capano}}, \bibinfo {author} {\bibfnamefont
  {I.}~\bibnamefont {Harry}}, \bibinfo {author} {\bibfnamefont
  {S.}~\bibnamefont {Mozzon}}, \bibinfo {author} {\bibfnamefont
  {L.}~\bibnamefont {Nuttall}}, \bibinfo {author} {\bibfnamefont
  {A.}~\bibnamefont {Lundgren}}, \ and\ \bibinfo {author} {\bibfnamefont
  {M.}~\bibnamefont {T\'apai}},\ }\href {\doibase 10.3847/1538-4357/ab733f}
  {\bibfield  {journal} {\bibinfo  {journal} {Astrophys. J.}\ }\textbf
  {\bibinfo {volume} {891}},\ \bibinfo {pages} {123} (\bibinfo {year}
  {2020})},\ \Eprint {http://arxiv.org/abs/1910.05331} {arXiv:1910.05331
  [astro-ph.HE]} \BibitemShut {NoStop}%
\bibitem [{\citenamefont {Venumadhav}\ \emph {et~al.}(2020)\citenamefont
  {Venumadhav}, \citenamefont {Zackay}, \citenamefont {Roulet}, \citenamefont
  {Dai},\ and\ \citenamefont {Zaldarriaga}}]{Venumadhav:2019lyq}%
  \BibitemOpen
  \bibfield  {author} {\bibinfo {author} {\bibfnamefont {T.}~\bibnamefont
  {Venumadhav}}, \bibinfo {author} {\bibfnamefont {B.}~\bibnamefont {Zackay}},
  \bibinfo {author} {\bibfnamefont {J.}~\bibnamefont {Roulet}}, \bibinfo
  {author} {\bibfnamefont {L.}~\bibnamefont {Dai}}, \ and\ \bibinfo {author}
  {\bibfnamefont {M.}~\bibnamefont {Zaldarriaga}},\ }\href {\doibase
  10.1103/PhysRevD.101.083030} {\bibfield  {journal} {\bibinfo  {journal}
  {Phys. Rev. D}\ }\textbf {\bibinfo {volume} {101}},\ \bibinfo {pages}
  {083030} (\bibinfo {year} {2020})},\ \Eprint
  {http://arxiv.org/abs/1904.07214} {arXiv:1904.07214 [astro-ph.HE]}
  \BibitemShut {NoStop}%
\bibitem [{\citenamefont {Zackay}\ \emph {et~al.}(2019)\citenamefont {Zackay},
  \citenamefont {Venumadhav}, \citenamefont {Dai}, \citenamefont {Roulet},\
  and\ \citenamefont {Zaldarriaga}}]{Zackay:2019tzo}%
  \BibitemOpen
  \bibfield  {author} {\bibinfo {author} {\bibfnamefont {B.}~\bibnamefont
  {Zackay}}, \bibinfo {author} {\bibfnamefont {T.}~\bibnamefont {Venumadhav}},
  \bibinfo {author} {\bibfnamefont {L.}~\bibnamefont {Dai}}, \bibinfo {author}
  {\bibfnamefont {J.}~\bibnamefont {Roulet}}, \ and\ \bibinfo {author}
  {\bibfnamefont {M.}~\bibnamefont {Zaldarriaga}},\ }\href {\doibase
  10.1103/PhysRevD.100.023007} {\bibfield  {journal} {\bibinfo  {journal}
  {Phys. Rev. D}\ }\textbf {\bibinfo {volume} {100}},\ \bibinfo {pages}
  {023007} (\bibinfo {year} {2019})},\ \Eprint
  {http://arxiv.org/abs/1902.10331} {arXiv:1902.10331 [astro-ph.HE]}
  \BibitemShut {NoStop}%
\bibitem [{\citenamefont {Peters}\ and\ \citenamefont
  {Mathews}(1963)}]{Peters:1963ux}%
  \BibitemOpen
  \bibfield  {author} {\bibinfo {author} {\bibfnamefont {P.~C.}\ \bibnamefont
  {Peters}}\ and\ \bibinfo {author} {\bibfnamefont {J.}~\bibnamefont
  {Mathews}},\ }\href {\doibase 10.1103/PhysRev.131.435} {\bibfield  {journal}
  {\bibinfo  {journal} {Phys. Rev.}\ }\textbf {\bibinfo {volume} {131}},\
  \bibinfo {pages} {435} (\bibinfo {year} {1963})}\BibitemShut {NoStop}%
\bibitem [{\citenamefont {Abbott}\ \emph
  {et~al.}(2017{\natexlab{a}})\citenamefont {Abbott} \emph
  {et~al.}}]{Abbott:2016bqf}%
  \BibitemOpen
  \bibfield  {author} {\bibinfo {author} {\bibfnamefont {B.~P.}\ \bibnamefont
  {Abbott}} \emph {et~al.} (\bibinfo {collaboration} {LIGO Scientific,
  Virgo}),\ }\href {\doibase 10.1002/andp.201600209} {\bibfield  {journal}
  {\bibinfo  {journal} {Annalen Phys.}\ }\textbf {\bibinfo {volume} {529}},\
  \bibinfo {pages} {1600209} (\bibinfo {year} {2017}{\natexlab{a}})},\ \Eprint
  {http://arxiv.org/abs/1608.01940} {arXiv:1608.01940 [gr-qc]} \BibitemShut
  {NoStop}%
\bibitem [{\citenamefont {Cutler}\ and\ \citenamefont
  {Flanagan}(1994)}]{Cutler:1994ys}%
  \BibitemOpen
  \bibfield  {author} {\bibinfo {author} {\bibfnamefont {C.}~\bibnamefont
  {Cutler}}\ and\ \bibinfo {author} {\bibfnamefont {E.~E.}\ \bibnamefont
  {Flanagan}},\ }\href {\doibase 10.1103/PhysRevD.49.2658} {\bibfield
  {journal} {\bibinfo  {journal} {Phys. Rev. D}\ }\textbf {\bibinfo {volume}
  {49}},\ \bibinfo {pages} {2658} (\bibinfo {year} {1994})},\ \Eprint
  {http://arxiv.org/abs/gr-qc/9402014} {arXiv:gr-qc/9402014} \BibitemShut
  {NoStop}%
\bibitem [{\citenamefont {Buonanno}\ and\ \citenamefont
  {Damour}(1999)}]{Buonanno:1998gg}%
  \BibitemOpen
  \bibfield  {author} {\bibinfo {author} {\bibfnamefont {A.}~\bibnamefont
  {Buonanno}}\ and\ \bibinfo {author} {\bibfnamefont {T.}~\bibnamefont
  {Damour}},\ }\href {\doibase 10.1103/PhysRevD.59.084006} {\bibfield
  {journal} {\bibinfo  {journal} {Phys. Rev. D}\ }\textbf {\bibinfo {volume}
  {59}},\ \bibinfo {pages} {084006} (\bibinfo {year} {1999})},\ \Eprint
  {http://arxiv.org/abs/gr-qc/9811091} {arXiv:gr-qc/9811091} \BibitemShut
  {NoStop}%
\bibitem [{\citenamefont {Damour}\ and\ \citenamefont
  {Nagar}(2016)}]{Damour:2016bks}%
  \BibitemOpen
  \bibfield  {author} {\bibinfo {author} {\bibfnamefont {T.}~\bibnamefont
  {Damour}}\ and\ \bibinfo {author} {\bibfnamefont {A.}~\bibnamefont {Nagar}},\
  }\enquote {\bibinfo {title} {{The Effective-One-Body Approach to the General
  Relativistic Two Body Problem}},}\ \ (\bibinfo {year} {2016})\ pp.\ \bibinfo
  {pages} {273--312}\BibitemShut {NoStop}%
\bibitem [{\citenamefont {Ossokine}\ \emph {et~al.}(2020)\citenamefont
  {Ossokine} \emph {et~al.}}]{Ossokine:2020kjp}%
  \BibitemOpen
  \bibfield  {author} {\bibinfo {author} {\bibfnamefont {S.}~\bibnamefont
  {Ossokine}} \emph {et~al.},\ }\href@noop {} {\  (\bibinfo {year} {2020})},\
  \Eprint {http://arxiv.org/abs/2004.09442} {arXiv:2004.09442 [gr-qc]}
  \BibitemShut {NoStop}%
\bibitem [{\citenamefont {Bohé}\ \emph {et~al.}(2017)\citenamefont {Bohé}
  \emph {et~al.}}]{Bohe:2016gbl}%
  \BibitemOpen
  \bibfield  {author} {\bibinfo {author} {\bibfnamefont {A.}~\bibnamefont
  {Bohé}} \emph {et~al.},\ }\href {\doibase 10.1103/PhysRevD.95.044028}
  {\bibfield  {journal} {\bibinfo  {journal} {Phys. Rev. D}\ }\textbf {\bibinfo
  {volume} {95}},\ \bibinfo {pages} {044028} (\bibinfo {year} {2017})},\
  \Eprint {http://arxiv.org/abs/1611.03703} {arXiv:1611.03703 [gr-qc]}
  \BibitemShut {NoStop}%
\bibitem [{\citenamefont {Taracchini}\ \emph {et~al.}(2014)\citenamefont
  {Taracchini} \emph {et~al.}}]{Taracchini:2013rva}%
  \BibitemOpen
  \bibfield  {author} {\bibinfo {author} {\bibfnamefont {A.}~\bibnamefont
  {Taracchini}} \emph {et~al.},\ }\href {\doibase 10.1103/PhysRevD.89.061502}
  {\bibfield  {journal} {\bibinfo  {journal} {Phys. Rev. D}\ }\textbf {\bibinfo
  {volume} {89}},\ \bibinfo {pages} {061502} (\bibinfo {year} {2014})},\
  \Eprint {http://arxiv.org/abs/1311.2544} {arXiv:1311.2544 [gr-qc]}
  \BibitemShut {NoStop}%
\bibitem [{\citenamefont {Pan}\ \emph {et~al.}(2014)\citenamefont {Pan},
  \citenamefont {Buonanno}, \citenamefont {Taracchini}, \citenamefont {Kidder},
  \citenamefont {Mroué}, \citenamefont {Pfeiffer}, \citenamefont {Scheel},\
  and\ \citenamefont {Szilágyi}}]{Pan:2013rra}%
  \BibitemOpen
  \bibfield  {author} {\bibinfo {author} {\bibfnamefont {Y.}~\bibnamefont
  {Pan}}, \bibinfo {author} {\bibfnamefont {A.}~\bibnamefont {Buonanno}},
  \bibinfo {author} {\bibfnamefont {A.}~\bibnamefont {Taracchini}}, \bibinfo
  {author} {\bibfnamefont {L.~E.}\ \bibnamefont {Kidder}}, \bibinfo {author}
  {\bibfnamefont {A.~H.}\ \bibnamefont {Mroué}}, \bibinfo {author}
  {\bibfnamefont {H.~P.}\ \bibnamefont {Pfeiffer}}, \bibinfo {author}
  {\bibfnamefont {M.~A.}\ \bibnamefont {Scheel}}, \ and\ \bibinfo {author}
  {\bibfnamefont {B.}~\bibnamefont {Szilágyi}},\ }\href {\doibase
  10.1103/PhysRevD.89.084006} {\bibfield  {journal} {\bibinfo  {journal} {Phys.
  Rev. D}\ }\textbf {\bibinfo {volume} {89}},\ \bibinfo {pages} {084006}
  (\bibinfo {year} {2014})},\ \Eprint {http://arxiv.org/abs/1307.6232}
  {arXiv:1307.6232 [gr-qc]} \BibitemShut {NoStop}%
\bibitem [{\citenamefont {Pan}\ \emph {et~al.}(2011)\citenamefont {Pan},
  \citenamefont {Buonanno}, \citenamefont {Boyle}, \citenamefont {Buchman},
  \citenamefont {Kidder}, \citenamefont {Pfeiffer},\ and\ \citenamefont
  {Scheel}}]{Pan:2011gk}%
  \BibitemOpen
  \bibfield  {author} {\bibinfo {author} {\bibfnamefont {Y.}~\bibnamefont
  {Pan}}, \bibinfo {author} {\bibfnamefont {A.}~\bibnamefont {Buonanno}},
  \bibinfo {author} {\bibfnamefont {M.}~\bibnamefont {Boyle}}, \bibinfo
  {author} {\bibfnamefont {L.~T.}\ \bibnamefont {Buchman}}, \bibinfo {author}
  {\bibfnamefont {L.~E.}\ \bibnamefont {Kidder}}, \bibinfo {author}
  {\bibfnamefont {H.~P.}\ \bibnamefont {Pfeiffer}}, \ and\ \bibinfo {author}
  {\bibfnamefont {M.~A.}\ \bibnamefont {Scheel}},\ }\href {\doibase
  10.1103/PhysRevD.84.124052} {\bibfield  {journal} {\bibinfo  {journal} {Phys.
  Rev. D}\ }\textbf {\bibinfo {volume} {84}},\ \bibinfo {pages} {124052}
  (\bibinfo {year} {2011})},\ \Eprint {http://arxiv.org/abs/1106.1021}
  {arXiv:1106.1021 [gr-qc]} \BibitemShut {NoStop}%
\bibitem [{\citenamefont {Nagar}\ \emph {et~al.}(2018)\citenamefont {Nagar}
  \emph {et~al.}}]{Nagar:2018zoe}%
  \BibitemOpen
  \bibfield  {author} {\bibinfo {author} {\bibfnamefont {A.}~\bibnamefont
  {Nagar}} \emph {et~al.},\ }\href {\doibase 10.1103/PhysRevD.98.104052}
  {\bibfield  {journal} {\bibinfo  {journal} {Phys. Rev. D}\ }\textbf {\bibinfo
  {volume} {98}},\ \bibinfo {pages} {104052} (\bibinfo {year} {2018})},\
  \Eprint {http://arxiv.org/abs/1806.01772} {arXiv:1806.01772 [gr-qc]}
  \BibitemShut {NoStop}%
\bibitem [{\citenamefont {Damour}\ and\ \citenamefont
  {Nagar}(2014)}]{Damour:2014sva}%
  \BibitemOpen
  \bibfield  {author} {\bibinfo {author} {\bibfnamefont {T.}~\bibnamefont
  {Damour}}\ and\ \bibinfo {author} {\bibfnamefont {A.}~\bibnamefont {Nagar}},\
  }\href {\doibase 10.1103/PhysRevD.90.044018} {\bibfield  {journal} {\bibinfo
  {journal} {Phys. Rev. D}\ }\textbf {\bibinfo {volume} {90}},\ \bibinfo
  {pages} {044018} (\bibinfo {year} {2014})},\ \Eprint
  {http://arxiv.org/abs/1406.6913} {arXiv:1406.6913 [gr-qc]} \BibitemShut
  {NoStop}%
\bibitem [{\citenamefont {Ajith}\ \emph {et~al.}(2008)\citenamefont {Ajith}
  \emph {et~al.}}]{Ajith:2007kx}%
  \BibitemOpen
  \bibfield  {author} {\bibinfo {author} {\bibfnamefont {P.}~\bibnamefont
  {Ajith}} \emph {et~al.},\ }\href {\doibase 10.1103/PhysRevD.77.104017}
  {\bibfield  {journal} {\bibinfo  {journal} {Phys. Rev. D}\ }\textbf {\bibinfo
  {volume} {77}},\ \bibinfo {pages} {104017} (\bibinfo {year} {2008})},\
  \bibinfo {note} {[Erratum: Phys.Rev.D 79, 129901 (2009)]},\ \Eprint
  {http://arxiv.org/abs/0710.2335} {arXiv:0710.2335 [gr-qc]} \BibitemShut
  {NoStop}%
\bibitem [{\citenamefont {Khan}\ \emph {et~al.}(2020)\citenamefont {Khan},
  \citenamefont {Ohme}, \citenamefont {Chatziioannou},\ and\ \citenamefont
  {Hannam}}]{Khan:2019kot}%
  \BibitemOpen
  \bibfield  {author} {\bibinfo {author} {\bibfnamefont {S.}~\bibnamefont
  {Khan}}, \bibinfo {author} {\bibfnamefont {F.}~\bibnamefont {Ohme}}, \bibinfo
  {author} {\bibfnamefont {K.}~\bibnamefont {Chatziioannou}}, \ and\ \bibinfo
  {author} {\bibfnamefont {M.}~\bibnamefont {Hannam}},\ }\href {\doibase
  10.1103/PhysRevD.101.024056} {\bibfield  {journal} {\bibinfo  {journal}
  {Phys. Rev. D}\ }\textbf {\bibinfo {volume} {101}},\ \bibinfo {pages}
  {024056} (\bibinfo {year} {2020})},\ \Eprint
  {http://arxiv.org/abs/1911.06050} {arXiv:1911.06050 [gr-qc]} \BibitemShut
  {NoStop}%
\bibitem [{\citenamefont {Khan}\ \emph {et~al.}(2019)\citenamefont {Khan},
  \citenamefont {Chatziioannou}, \citenamefont {Hannam},\ and\ \citenamefont
  {Ohme}}]{Khan:2018fmp}%
  \BibitemOpen
  \bibfield  {author} {\bibinfo {author} {\bibfnamefont {S.}~\bibnamefont
  {Khan}}, \bibinfo {author} {\bibfnamefont {K.}~\bibnamefont {Chatziioannou}},
  \bibinfo {author} {\bibfnamefont {M.}~\bibnamefont {Hannam}}, \ and\ \bibinfo
  {author} {\bibfnamefont {F.}~\bibnamefont {Ohme}},\ }\href {\doibase
  10.1103/PhysRevD.100.024059} {\bibfield  {journal} {\bibinfo  {journal}
  {Phys. Rev. D}\ }\textbf {\bibinfo {volume} {100}},\ \bibinfo {pages}
  {024059} (\bibinfo {year} {2019})},\ \Eprint
  {http://arxiv.org/abs/1809.10113} {arXiv:1809.10113 [gr-qc]} \BibitemShut
  {NoStop}%
\bibitem [{\citenamefont {London}\ \emph {et~al.}(2018)\citenamefont {London},
  \citenamefont {Khan}, \citenamefont {Fauchon-Jones}, \citenamefont {García},
  \citenamefont {Hannam}, \citenamefont {Husa}, \citenamefont
  {Jiménez-Forteza}, \citenamefont {Kalaghatgi}, \citenamefont {Ohme},\ and\
  \citenamefont {Pannarale}}]{London:2017bcn}%
  \BibitemOpen
  \bibfield  {author} {\bibinfo {author} {\bibfnamefont {L.}~\bibnamefont
  {London}}, \bibinfo {author} {\bibfnamefont {S.}~\bibnamefont {Khan}},
  \bibinfo {author} {\bibfnamefont {E.}~\bibnamefont {Fauchon-Jones}}, \bibinfo
  {author} {\bibfnamefont {C.}~\bibnamefont {García}}, \bibinfo {author}
  {\bibfnamefont {M.}~\bibnamefont {Hannam}}, \bibinfo {author} {\bibfnamefont
  {S.}~\bibnamefont {Husa}}, \bibinfo {author} {\bibfnamefont {X.}~\bibnamefont
  {Jiménez-Forteza}}, \bibinfo {author} {\bibfnamefont {C.}~\bibnamefont
  {Kalaghatgi}}, \bibinfo {author} {\bibfnamefont {F.}~\bibnamefont {Ohme}}, \
  and\ \bibinfo {author} {\bibfnamefont {F.}~\bibnamefont {Pannarale}},\ }\href
  {\doibase 10.1103/PhysRevLett.120.161102} {\bibfield  {journal} {\bibinfo
  {journal} {Phys. Rev. Lett.}\ }\textbf {\bibinfo {volume} {120}},\ \bibinfo
  {pages} {161102} (\bibinfo {year} {2018})},\ \Eprint
  {http://arxiv.org/abs/1708.00404} {arXiv:1708.00404 [gr-qc]} \BibitemShut
  {NoStop}%
\bibitem [{\citenamefont {Khan}\ \emph {et~al.}(2016)\citenamefont {Khan},
  \citenamefont {Husa}, \citenamefont {Hannam}, \citenamefont {Ohme},
  \citenamefont {Pürrer}, \citenamefont {Jiménez~Forteza},\ and\
  \citenamefont {Bohé}}]{Khan:2015jqa}%
  \BibitemOpen
  \bibfield  {author} {\bibinfo {author} {\bibfnamefont {S.}~\bibnamefont
  {Khan}}, \bibinfo {author} {\bibfnamefont {S.}~\bibnamefont {Husa}}, \bibinfo
  {author} {\bibfnamefont {M.}~\bibnamefont {Hannam}}, \bibinfo {author}
  {\bibfnamefont {F.}~\bibnamefont {Ohme}}, \bibinfo {author} {\bibfnamefont
  {M.}~\bibnamefont {Pürrer}}, \bibinfo {author} {\bibfnamefont
  {X.}~\bibnamefont {Jiménez~Forteza}}, \ and\ \bibinfo {author}
  {\bibfnamefont {A.}~\bibnamefont {Bohé}},\ }\href {\doibase
  10.1103/PhysRevD.93.044007} {\bibfield  {journal} {\bibinfo  {journal} {Phys.
  Rev. D}\ }\textbf {\bibinfo {volume} {93}},\ \bibinfo {pages} {044007}
  (\bibinfo {year} {2016})},\ \Eprint {http://arxiv.org/abs/1508.07253}
  {arXiv:1508.07253 [gr-qc]} \BibitemShut {NoStop}%
\bibitem [{\citenamefont {Husa}\ \emph {et~al.}(2016)\citenamefont {Husa},
  \citenamefont {Khan}, \citenamefont {Hannam}, \citenamefont {Pürrer},
  \citenamefont {Ohme}, \citenamefont {Jiménez~Forteza},\ and\ \citenamefont
  {Bohé}}]{Husa:2015iqa}%
  \BibitemOpen
  \bibfield  {author} {\bibinfo {author} {\bibfnamefont {S.}~\bibnamefont
  {Husa}}, \bibinfo {author} {\bibfnamefont {S.}~\bibnamefont {Khan}}, \bibinfo
  {author} {\bibfnamefont {M.}~\bibnamefont {Hannam}}, \bibinfo {author}
  {\bibfnamefont {M.}~\bibnamefont {Pürrer}}, \bibinfo {author} {\bibfnamefont
  {F.}~\bibnamefont {Ohme}}, \bibinfo {author} {\bibfnamefont {X.}~\bibnamefont
  {Jiménez~Forteza}}, \ and\ \bibinfo {author} {\bibfnamefont
  {A.}~\bibnamefont {Bohé}},\ }\href {\doibase 10.1103/PhysRevD.93.044006}
  {\bibfield  {journal} {\bibinfo  {journal} {Phys. Rev. D}\ }\textbf {\bibinfo
  {volume} {93}},\ \bibinfo {pages} {044006} (\bibinfo {year} {2016})},\
  \Eprint {http://arxiv.org/abs/1508.07250} {arXiv:1508.07250 [gr-qc]}
  \BibitemShut {NoStop}%
\bibitem [{\citenamefont {Santamaria}\ \emph {et~al.}(2010)\citenamefont
  {Santamaria} \emph {et~al.}}]{Santamaria:2010yb}%
  \BibitemOpen
  \bibfield  {author} {\bibinfo {author} {\bibfnamefont {L.}~\bibnamefont
  {Santamaria}} \emph {et~al.},\ }\href {\doibase 10.1103/PhysRevD.82.064016}
  {\bibfield  {journal} {\bibinfo  {journal} {Phys. Rev. D}\ }\textbf {\bibinfo
  {volume} {82}},\ \bibinfo {pages} {064016} (\bibinfo {year} {2010})},\
  \Eprint {http://arxiv.org/abs/1005.3306} {arXiv:1005.3306 [gr-qc]}
  \BibitemShut {NoStop}%
\bibitem [{\citenamefont {Field}\ \emph {et~al.}(2014)\citenamefont {Field},
  \citenamefont {Galley}, \citenamefont {Hesthaven}, \citenamefont {Kaye},\
  and\ \citenamefont {Tiglio}}]{Field:2013cfa}%
  \BibitemOpen
  \bibfield  {author} {\bibinfo {author} {\bibfnamefont {S.~E.}\ \bibnamefont
  {Field}}, \bibinfo {author} {\bibfnamefont {C.~R.}\ \bibnamefont {Galley}},
  \bibinfo {author} {\bibfnamefont {J.~S.}\ \bibnamefont {Hesthaven}}, \bibinfo
  {author} {\bibfnamefont {J.}~\bibnamefont {Kaye}}, \ and\ \bibinfo {author}
  {\bibfnamefont {M.}~\bibnamefont {Tiglio}},\ }\href {\doibase
  10.1103/PhysRevX.4.031006} {\bibfield  {journal} {\bibinfo  {journal} {Phys.
  Rev. X}\ }\textbf {\bibinfo {volume} {4}},\ \bibinfo {pages} {031006}
  (\bibinfo {year} {2014})},\ \Eprint {http://arxiv.org/abs/1308.3565}
  {arXiv:1308.3565 [gr-qc]} \BibitemShut {NoStop}%
\bibitem [{\citenamefont {Blackman}\ \emph {et~al.}(2017)\citenamefont
  {Blackman}, \citenamefont {Field}, \citenamefont {Scheel}, \citenamefont
  {Galley}, \citenamefont {Ott}, \citenamefont {Boyle}, \citenamefont {Kidder},
  \citenamefont {Pfeiffer},\ and\ \citenamefont
  {Szil\'agyi}}]{Blackman:2017pcm}%
  \BibitemOpen
  \bibfield  {author} {\bibinfo {author} {\bibfnamefont {J.}~\bibnamefont
  {Blackman}}, \bibinfo {author} {\bibfnamefont {S.~E.}\ \bibnamefont {Field}},
  \bibinfo {author} {\bibfnamefont {M.~A.}\ \bibnamefont {Scheel}}, \bibinfo
  {author} {\bibfnamefont {C.~R.}\ \bibnamefont {Galley}}, \bibinfo {author}
  {\bibfnamefont {C.~D.}\ \bibnamefont {Ott}}, \bibinfo {author} {\bibfnamefont
  {M.}~\bibnamefont {Boyle}}, \bibinfo {author} {\bibfnamefont {L.~E.}\
  \bibnamefont {Kidder}}, \bibinfo {author} {\bibfnamefont {H.~P.}\
  \bibnamefont {Pfeiffer}}, \ and\ \bibinfo {author} {\bibfnamefont
  {B.}~\bibnamefont {Szil\'agyi}},\ }\href {\doibase
  10.1103/PhysRevD.96.024058} {\bibfield  {journal} {\bibinfo  {journal} {Phys.
  Rev. D}\ }\textbf {\bibinfo {volume} {96}},\ \bibinfo {pages} {024058}
  (\bibinfo {year} {2017})},\ \Eprint {http://arxiv.org/abs/1705.07089}
  {arXiv:1705.07089 [gr-qc]} \BibitemShut {NoStop}%
\bibitem [{\citenamefont {Varma}\ \emph {et~al.}(2019)\citenamefont {Varma},
  \citenamefont {Field}, \citenamefont {Scheel}, \citenamefont {Blackman},
  \citenamefont {Gerosa}, \citenamefont {Stein}, \citenamefont {Kidder},\ and\
  \citenamefont {Pfeiffer}}]{PhysRevResearch.1.033015}%
  \BibitemOpen
  \bibfield  {author} {\bibinfo {author} {\bibfnamefont {V.}~\bibnamefont
  {Varma}}, \bibinfo {author} {\bibfnamefont {S.~E.}\ \bibnamefont {Field}},
  \bibinfo {author} {\bibfnamefont {M.~A.}\ \bibnamefont {Scheel}}, \bibinfo
  {author} {\bibfnamefont {J.}~\bibnamefont {Blackman}}, \bibinfo {author}
  {\bibfnamefont {D.}~\bibnamefont {Gerosa}}, \bibinfo {author} {\bibfnamefont
  {L.~C.}\ \bibnamefont {Stein}}, \bibinfo {author} {\bibfnamefont {L.~E.}\
  \bibnamefont {Kidder}}, \ and\ \bibinfo {author} {\bibfnamefont {H.~P.}\
  \bibnamefont {Pfeiffer}},\ }\href {\doibase 10.1103/PhysRevResearch.1.033015}
  {\bibfield  {journal} {\bibinfo  {journal} {Phys. Rev. Research}\ }\textbf
  {\bibinfo {volume} {1}},\ \bibinfo {pages} {033015} (\bibinfo {year}
  {2019})}\BibitemShut {NoStop}%
\bibitem [{\citenamefont {Ohashi}\ \emph {et~al.}(1996)\citenamefont {Ohashi},
  \citenamefont {Tagoshi},\ and\ \citenamefont {Sasaki}}]{Ohashi:1996uz}%
  \BibitemOpen
  \bibfield  {author} {\bibinfo {author} {\bibfnamefont {A.}~\bibnamefont
  {Ohashi}}, \bibinfo {author} {\bibfnamefont {H.}~\bibnamefont {Tagoshi}}, \
  and\ \bibinfo {author} {\bibfnamefont {M.}~\bibnamefont {Sasaki}},\ }\href
  {\doibase 10.1143/PTP.96.713} {\bibfield  {journal} {\bibinfo  {journal}
  {Prog. Theor. Phys.}\ }\textbf {\bibinfo {volume} {96}},\ \bibinfo {pages}
  {713} (\bibinfo {year} {1996})}\BibitemShut {NoStop}%
\bibitem [{\citenamefont {Nielsen}\ and\ \citenamefont
  {Birnholtz}(2018)}]{Nielsen:2017lpd}%
  \BibitemOpen
  \bibfield  {author} {\bibinfo {author} {\bibfnamefont {A.~B.}\ \bibnamefont
  {Nielsen}}\ and\ \bibinfo {author} {\bibfnamefont {O.}~\bibnamefont
  {Birnholtz}},\ }\href {\doibase 10.1002/asna.201813473} {\bibfield  {journal}
  {\bibinfo  {journal} {Astron. Nachr.}\ }\textbf {\bibinfo {volume} {339}},\
  \bibinfo {pages} {298} (\bibinfo {year} {2018})},\ \Eprint
  {http://arxiv.org/abs/1708.03334} {arXiv:1708.03334 [gr-qc]} \BibitemShut
  {NoStop}%
\bibitem [{\citenamefont {Abbott}\ \emph
  {et~al.}(2017{\natexlab{b}})\citenamefont {Abbott} \emph
  {et~al.}}]{TheLIGOScientific:2017qsa}%
  \BibitemOpen
  \bibfield  {author} {\bibinfo {author} {\bibfnamefont {B.~P.}\ \bibnamefont
  {Abbott}} \emph {et~al.} (\bibinfo {collaboration} {LIGO Scientific,
  Virgo}),\ }\href {\doibase 10.1103/PhysRevLett.119.161101} {\bibfield
  {journal} {\bibinfo  {journal} {Phys. Rev. Lett.}\ }\textbf {\bibinfo
  {volume} {119}},\ \bibinfo {pages} {161101} (\bibinfo {year}
  {2017}{\natexlab{b}})},\ \Eprint {http://arxiv.org/abs/1710.05832}
  {arXiv:1710.05832 [gr-qc]} \BibitemShut {NoStop}%
\bibitem [{\citenamefont {Arun}\ \emph
  {et~al.}(2006{\natexlab{a}})\citenamefont {Arun}, \citenamefont {Iyer},
  \citenamefont {Qusailah},\ and\ \citenamefont {Sathyaprakash}}]{Arun:2006yw}%
  \BibitemOpen
  \bibfield  {author} {\bibinfo {author} {\bibfnamefont {K.~G.}\ \bibnamefont
  {Arun}}, \bibinfo {author} {\bibfnamefont {B.~R.}\ \bibnamefont {Iyer}},
  \bibinfo {author} {\bibfnamefont {M.~S.~S.}\ \bibnamefont {Qusailah}}, \ and\
  \bibinfo {author} {\bibfnamefont {B.~S.}\ \bibnamefont {Sathyaprakash}},\
  }\href {\doibase 10.1088/0264-9381/23/9/L01} {\bibfield  {journal} {\bibinfo
  {journal} {Class. Quant. Grav.}\ }\textbf {\bibinfo {volume} {23}},\ \bibinfo
  {pages} {L37} (\bibinfo {year} {2006}{\natexlab{a}})},\ \Eprint
  {http://arxiv.org/abs/gr-qc/0604018} {arXiv:gr-qc/0604018} \BibitemShut
  {NoStop}%
\bibitem [{\citenamefont {Arun}\ \emph
  {et~al.}(2006{\natexlab{b}})\citenamefont {Arun}, \citenamefont {Iyer},
  \citenamefont {Qusailah},\ and\ \citenamefont {Sathyaprakash}}]{Arun:2006hn}%
  \BibitemOpen
  \bibfield  {author} {\bibinfo {author} {\bibfnamefont {K.~G.}\ \bibnamefont
  {Arun}}, \bibinfo {author} {\bibfnamefont {B.~R.}\ \bibnamefont {Iyer}},
  \bibinfo {author} {\bibfnamefont {M.~S.~S.}\ \bibnamefont {Qusailah}}, \ and\
  \bibinfo {author} {\bibfnamefont {B.~S.}\ \bibnamefont {Sathyaprakash}},\
  }\href {\doibase 10.1103/PhysRevD.74.024006} {\bibfield  {journal} {\bibinfo
  {journal} {Phys. Rev. D}\ }\textbf {\bibinfo {volume} {74}},\ \bibinfo
  {pages} {024006} (\bibinfo {year} {2006}{\natexlab{b}})},\ \Eprint
  {http://arxiv.org/abs/gr-qc/0604067} {arXiv:gr-qc/0604067} \BibitemShut
  {NoStop}%
\bibitem [{\citenamefont {Mishra}\ \emph {et~al.}(2010)\citenamefont {Mishra},
  \citenamefont {Arun}, \citenamefont {Iyer},\ and\ \citenamefont
  {Sathyaprakash}}]{Mishra:2010tp}%
  \BibitemOpen
  \bibfield  {author} {\bibinfo {author} {\bibfnamefont {C.~K.}\ \bibnamefont
  {Mishra}}, \bibinfo {author} {\bibfnamefont {K.~G.}\ \bibnamefont {Arun}},
  \bibinfo {author} {\bibfnamefont {B.~R.}\ \bibnamefont {Iyer}}, \ and\
  \bibinfo {author} {\bibfnamefont {B.~S.}\ \bibnamefont {Sathyaprakash}},\
  }\href {\doibase 10.1103/PhysRevD.82.064010} {\bibfield  {journal} {\bibinfo
  {journal} {Phys. Rev. D}\ }\textbf {\bibinfo {volume} {82}},\ \bibinfo
  {pages} {064010} (\bibinfo {year} {2010})},\ \Eprint
  {http://arxiv.org/abs/1005.0304} {arXiv:1005.0304 [gr-qc]} \BibitemShut
  {NoStop}%
\bibitem [{\citenamefont {Abbott}\ \emph
  {et~al.}(2016{\natexlab{c}})\citenamefont {Abbott} \emph
  {et~al.}}]{TheLIGOScientific:2016src}%
  \BibitemOpen
  \bibfield  {author} {\bibinfo {author} {\bibfnamefont {B.~P.}\ \bibnamefont
  {Abbott}} \emph {et~al.} (\bibinfo {collaboration} {LIGO Scientific,
  Virgo}),\ }\href {\doibase 10.1103/PhysRevLett.116.221101} {\bibfield
  {journal} {\bibinfo  {journal} {Phys. Rev. Lett.}\ }\textbf {\bibinfo
  {volume} {116}},\ \bibinfo {pages} {221101} (\bibinfo {year}
  {2016}{\natexlab{c}})},\ \bibinfo {note} {[Erratum: Phys.Rev.Lett. 121,
  129902 (2018)]},\ \Eprint {http://arxiv.org/abs/1602.03841} {arXiv:1602.03841
  [gr-qc]} \BibitemShut {NoStop}%
\bibitem [{\citenamefont {Abbott}\ \emph
  {et~al.}(2019{\natexlab{b}})\citenamefont {Abbott} \emph
  {et~al.}}]{Abbott:2018lct}%
  \BibitemOpen
  \bibfield  {author} {\bibinfo {author} {\bibfnamefont {B.~P.}\ \bibnamefont
  {Abbott}} \emph {et~al.} (\bibinfo {collaboration} {LIGO Scientific,
  Virgo}),\ }\href {\doibase 10.1103/PhysRevLett.123.011102} {\bibfield
  {journal} {\bibinfo  {journal} {Phys. Rev. Lett.}\ }\textbf {\bibinfo
  {volume} {123}},\ \bibinfo {pages} {011102} (\bibinfo {year}
  {2019}{\natexlab{b}})},\ \Eprint {http://arxiv.org/abs/1811.00364}
  {arXiv:1811.00364 [gr-qc]} \BibitemShut {NoStop}%
\bibitem [{\citenamefont {Abbott}\ \emph
  {et~al.}(2019{\natexlab{c}})\citenamefont {Abbott} \emph
  {et~al.}}]{LIGOScientific:2019fpa}%
  \BibitemOpen
  \bibfield  {author} {\bibinfo {author} {\bibfnamefont {B.~P.}\ \bibnamefont
  {Abbott}} \emph {et~al.} (\bibinfo {collaboration} {LIGO Scientific,
  Virgo}),\ }\href {\doibase 10.1103/PhysRevD.100.104036} {\bibfield  {journal}
  {\bibinfo  {journal} {Phys. Rev. D}\ }\textbf {\bibinfo {volume} {100}},\
  \bibinfo {pages} {104036} (\bibinfo {year} {2019}{\natexlab{c}})},\ \Eprint
  {http://arxiv.org/abs/1903.04467} {arXiv:1903.04467 [gr-qc]} \BibitemShut
  {NoStop}%
\bibitem [{\citenamefont {Agathos}\ \emph {et~al.}(2014)\citenamefont
  {Agathos}, \citenamefont {Del~Pozzo}, \citenamefont {Li}, \citenamefont {Van
  Den~Broeck}, \citenamefont {Veitch},\ and\ \citenamefont
  {Vitale}}]{Agathos:2013upa}%
  \BibitemOpen
  \bibfield  {author} {\bibinfo {author} {\bibfnamefont {M.}~\bibnamefont
  {Agathos}}, \bibinfo {author} {\bibfnamefont {W.}~\bibnamefont {Del~Pozzo}},
  \bibinfo {author} {\bibfnamefont {T.~G.~F.}\ \bibnamefont {Li}}, \bibinfo
  {author} {\bibfnamefont {C.}~\bibnamefont {Van Den~Broeck}}, \bibinfo
  {author} {\bibfnamefont {J.}~\bibnamefont {Veitch}}, \ and\ \bibinfo {author}
  {\bibfnamefont {S.}~\bibnamefont {Vitale}},\ }\href {\doibase
  10.1103/PhysRevD.89.082001} {\bibfield  {journal} {\bibinfo  {journal} {Phys.
  Rev. D}\ }\textbf {\bibinfo {volume} {89}},\ \bibinfo {pages} {082001}
  (\bibinfo {year} {2014})},\ \Eprint {http://arxiv.org/abs/1311.0420}
  {arXiv:1311.0420 [gr-qc]} \BibitemShut {NoStop}%
\bibitem [{\citenamefont {Will}(1998)}]{Will:1997bb}%
  \BibitemOpen
  \bibfield  {author} {\bibinfo {author} {\bibfnamefont {C.~M.}\ \bibnamefont
  {Will}},\ }\href {\doibase 10.1103/PhysRevD.57.2061} {\bibfield  {journal}
  {\bibinfo  {journal} {Phys. Rev. D}\ }\textbf {\bibinfo {volume} {57}},\
  \bibinfo {pages} {2061} (\bibinfo {year} {1998})},\ \Eprint
  {http://arxiv.org/abs/gr-qc/9709011} {arXiv:gr-qc/9709011} \BibitemShut
  {NoStop}%
\bibitem [{\citenamefont {Kastha}\ \emph {et~al.}(2018)\citenamefont {Kastha},
  \citenamefont {Gupta}, \citenamefont {Arun}, \citenamefont {Sathyaprakash},\
  and\ \citenamefont {Van Den~Broeck}}]{Kastha:2018bcr}%
  \BibitemOpen
  \bibfield  {author} {\bibinfo {author} {\bibfnamefont {S.}~\bibnamefont
  {Kastha}}, \bibinfo {author} {\bibfnamefont {A.}~\bibnamefont {Gupta}},
  \bibinfo {author} {\bibfnamefont {K.~G.}\ \bibnamefont {Arun}}, \bibinfo
  {author} {\bibfnamefont {B.~S.}\ \bibnamefont {Sathyaprakash}}, \ and\
  \bibinfo {author} {\bibfnamefont {C.}~\bibnamefont {Van Den~Broeck}},\ }\href
  {\doibase 10.1103/PhysRevD.98.124033} {\bibfield  {journal} {\bibinfo
  {journal} {Phys. Rev. D}\ }\textbf {\bibinfo {volume} {98}},\ \bibinfo
  {pages} {124033} (\bibinfo {year} {2018})},\ \Eprint
  {http://arxiv.org/abs/1809.10465} {arXiv:1809.10465 [gr-qc]} \BibitemShut
  {NoStop}%
\bibitem [{\citenamefont {Kastha}\ \emph {et~al.}(2019)\citenamefont {Kastha},
  \citenamefont {Gupta}, \citenamefont {Arun}, \citenamefont {Sathyaprakash},\
  and\ \citenamefont {Van Den~Broeck}}]{Kastha:2019brk}%
  \BibitemOpen
  \bibfield  {author} {\bibinfo {author} {\bibfnamefont {S.}~\bibnamefont
  {Kastha}}, \bibinfo {author} {\bibfnamefont {A.}~\bibnamefont {Gupta}},
  \bibinfo {author} {\bibfnamefont {K.~G.}\ \bibnamefont {Arun}}, \bibinfo
  {author} {\bibfnamefont {B.~S.}\ \bibnamefont {Sathyaprakash}}, \ and\
  \bibinfo {author} {\bibfnamefont {C.}~\bibnamefont {Van Den~Broeck}},\ }\href
  {\doibase 10.1103/PhysRevD.100.044007} {\bibfield  {journal} {\bibinfo
  {journal} {Phys. Rev. D}\ }\textbf {\bibinfo {volume} {100}},\ \bibinfo
  {pages} {044007} (\bibinfo {year} {2019})},\ \Eprint
  {http://arxiv.org/abs/1905.07277} {arXiv:1905.07277 [gr-qc]} \BibitemShut
  {NoStop}%
\bibitem [{\citenamefont {Ohme}\ \emph {et~al.}(2013)\citenamefont {Ohme},
  \citenamefont {Nielsen}, \citenamefont {Keppel},\ and\ \citenamefont
  {Lundgren}}]{Ohme:2013nsa}%
  \BibitemOpen
  \bibfield  {author} {\bibinfo {author} {\bibfnamefont {F.}~\bibnamefont
  {Ohme}}, \bibinfo {author} {\bibfnamefont {A.~B.}\ \bibnamefont {Nielsen}},
  \bibinfo {author} {\bibfnamefont {D.}~\bibnamefont {Keppel}}, \ and\ \bibinfo
  {author} {\bibfnamefont {A.}~\bibnamefont {Lundgren}},\ }\href {\doibase
  10.1103/PhysRevD.88.042002} {\bibfield  {journal} {\bibinfo  {journal} {Phys.
  Rev. D}\ }\textbf {\bibinfo {volume} {88}},\ \bibinfo {pages} {042002}
  (\bibinfo {year} {2013})},\ \Eprint {http://arxiv.org/abs/1304.7017}
  {arXiv:1304.7017 [gr-qc]} \BibitemShut {NoStop}%
\bibitem [{\citenamefont {Pai}\ and\ \citenamefont {Arun}(2013)}]{Pai:2012mv}%
  \BibitemOpen
  \bibfield  {author} {\bibinfo {author} {\bibfnamefont {A.}~\bibnamefont
  {Pai}}\ and\ \bibinfo {author} {\bibfnamefont {K.~G.}\ \bibnamefont {Arun}},\
  }\href {\doibase 10.1088/0264-9381/30/2/025011} {\bibfield  {journal}
  {\bibinfo  {journal} {Class. Quant. Grav.}\ }\textbf {\bibinfo {volume}
  {30}},\ \bibinfo {pages} {025011} (\bibinfo {year} {2013})},\ \Eprint
  {http://arxiv.org/abs/1207.1943} {arXiv:1207.1943 [gr-qc]} \BibitemShut
  {NoStop}%
\bibitem [{\citenamefont {Vallisneri}(2008)}]{Vallisneri:2007ev}%
  \BibitemOpen
  \bibfield  {author} {\bibinfo {author} {\bibfnamefont {M.}~\bibnamefont
  {Vallisneri}},\ }\href {\doibase 10.1103/PhysRevD.77.042001} {\bibfield
  {journal} {\bibinfo  {journal} {Phys. Rev. D}\ }\textbf {\bibinfo {volume}
  {77}},\ \bibinfo {pages} {042001} (\bibinfo {year} {2008})},\ \Eprint
  {http://arxiv.org/abs/gr-qc/0703086} {arXiv:gr-qc/0703086} \BibitemShut
  {NoStop}%
\bibitem [{\citenamefont {Biwer}\ \emph {et~al.}(2019)\citenamefont {Biwer},
  \citenamefont {Capano}, \citenamefont {De}, \citenamefont {Cabero},
  \citenamefont {Brown}, \citenamefont {Nitz},\ and\ \citenamefont
  {Raymond}}]{Biwer:2018osg}%
  \BibitemOpen
  \bibfield  {author} {\bibinfo {author} {\bibfnamefont {C.~M.}\ \bibnamefont
  {Biwer}}, \bibinfo {author} {\bibfnamefont {C.~D.}\ \bibnamefont {Capano}},
  \bibinfo {author} {\bibfnamefont {S.}~\bibnamefont {De}}, \bibinfo {author}
  {\bibfnamefont {M.}~\bibnamefont {Cabero}}, \bibinfo {author} {\bibfnamefont
  {D.~A.}\ \bibnamefont {Brown}}, \bibinfo {author} {\bibfnamefont {A.~H.}\
  \bibnamefont {Nitz}}, \ and\ \bibinfo {author} {\bibfnamefont
  {V.}~\bibnamefont {Raymond}},\ }\href {\doibase 10.1088/1538-3873/aaef0b}
  {\bibfield  {journal} {\bibinfo  {journal} {Publ. Astron. Soc. Pac.}\
  }\textbf {\bibinfo {volume} {131}},\ \bibinfo {pages} {024503} (\bibinfo
  {year} {2019})},\ \Eprint {http://arxiv.org/abs/1807.10312} {arXiv:1807.10312
  [astro-ph.IM]} \BibitemShut {NoStop}%
\bibitem [{\citenamefont {Buonanno}\ \emph {et~al.}(2009)\citenamefont
  {Buonanno}, \citenamefont {Iyer}, \citenamefont {Ochsner}, \citenamefont
  {Pan},\ and\ \citenamefont {Sathyaprakash}}]{Buonanno:2009zt}%
  \BibitemOpen
  \bibfield  {author} {\bibinfo {author} {\bibfnamefont {A.}~\bibnamefont
  {Buonanno}}, \bibinfo {author} {\bibfnamefont {B.}~\bibnamefont {Iyer}},
  \bibinfo {author} {\bibfnamefont {E.}~\bibnamefont {Ochsner}}, \bibinfo
  {author} {\bibfnamefont {Y.}~\bibnamefont {Pan}}, \ and\ \bibinfo {author}
  {\bibfnamefont {B.~S.}\ \bibnamefont {Sathyaprakash}},\ }\href {\doibase
  10.1103/PhysRevD.80.084043} {\bibfield  {journal} {\bibinfo  {journal} {Phys.
  Rev. D}\ }\textbf {\bibinfo {volume} {80}},\ \bibinfo {pages} {084043}
  (\bibinfo {year} {2009})},\ \Eprint {http://arxiv.org/abs/0907.0700}
  {arXiv:0907.0700 [gr-qc]} \BibitemShut {NoStop}%
\bibitem [{\citenamefont {{LIGO Scientific Collaboration}}(2018)}]{lalsuite}%
  \BibitemOpen
  \bibfield  {author} {\bibinfo {author} {\bibnamefont {{LIGO Scientific
  Collaboration}}},\ }\href {\doibase 10.7935/GT1W-FZ16} {\enquote {\bibinfo
  {title} {{LIGO} {A}lgorithm {L}ibrary - {LALS}uite},}\ }\bibinfo
  {howpublished} {free software (GPL)} (\bibinfo {year} {2018})\BibitemShut
  {NoStop}%
\bibitem [{\citenamefont {Press}\ \emph {et~al.}(1992)\citenamefont {Press},
  \citenamefont {Teukolsky}, \citenamefont {Vetterling},\ and\ \citenamefont
  {Flannery}}]{Fortran}%
  \BibitemOpen
  \bibfield  {author} {\bibinfo {author} {\bibfnamefont {W.~H.}\ \bibnamefont
  {Press}}, \bibinfo {author} {\bibfnamefont {S.~A.}\ \bibnamefont
  {Teukolsky}}, \bibinfo {author} {\bibfnamefont {W.~T.}\ \bibnamefont
  {Vetterling}}, \ and\ \bibinfo {author} {\bibfnamefont {B.~P.}\ \bibnamefont
  {Flannery}},\ }\href@noop {} {\  (\bibinfo {year} {1992})}\BibitemShut
  {NoStop}%
\bibitem [{\citenamefont {AEI}(2017)}]{Atlas}%
  \BibitemOpen
  \bibfield  {author} {\bibinfo {author} {\bibnamefont {AEI}},\ }\href
  {https://www.aei.mpg.de/24838/02\_Computing\_and\_ATLAS} {\enquote {\bibinfo
  {title} {{The Atlas Computing Cluster}},}\ }\bibinfo {howpublished}
  {https://www.aei.mpg.de/24838/02\_Computing\_and\_ATLAS} (\bibinfo {year}
  {2017})\BibitemShut {NoStop}%
\end{thebibliography}%

\end{document}